\documentclass[12pt,preprint]{aastex}
\usepackage{emulateapj5}

\slugcomment{Not to appear in Nonlearned J., 45.}

\shorttitle{The Kramers-Heisenberg Formula and the Gunn-Peterson
Troughs}

\shortauthors{Bach \& Lee}

\begin{document}

\title{The Kramers-Heisenberg Formula and the Gunn-Peterson Trough
from the First Objects in the Universe}

\author{Kiehunn Bach}
\affil{Department of Astronomy, Yonsei University,
    Seoul, Korea}

\and

\author{Hee-Won Lee}
\affil{Department of Earth Sciences, Sejong University, Seoul,
Korea \\
{\rm hwlee@sejong.ac.kr}}

\begin{abstract}
Recent cosmological observations indicate that the
reionized universe may have started at around $z=6$, where a
significant suppression around Ly$\alpha$ has been observed from
the neutral intergalactic medium. The associated neutral
hydrogen column density is expected to exceed $10^{21}{\rm\ cm^{-2}}$,
where it is very important to use the accurate
scattering cross section known as the Kramers-Heisenberg formula that is
obtained from the fully quantum mechanical time-dependent second order
perturbation theory. We present the 
Kramers-Heisenberg formula and compare
it with the formula introduced in a heuristic way by Peebles (1993)
treating the hydrogen atom as a two-level atom, from which we find a devitaion
by a factor of two in the red wing region far from the line center.
Adopting simple cosmological models, we compute the Gunn-Peterson 
optical depths and the trough profiles.
Our results are compared with the works performed by Madau \& Rees
(2000), who adopted the cross section
introduced by Peebles (1993). We find deviations up to 5 per cent in the
Gunn-Peterson transmission coefficient for an accelerated expanding universe 
in the red off-resonance 
wing part with the rest wavelength $\Delta\lambda\sim 10{\rm\ \AA}$.
\end{abstract}


\section{Introduction}

The quasar absorption systems have been excellent tools to
investigate the intergalactic medium (IGM), from which
it has been well known that the IGM of the nearby universe is highly
ionized (e.g. Peebles 1993). Since the universe after the
recombination era $z\sim 1100$ should be dominantly neutral, there
must be some epoch when the universe began to be re-ionized.
Intensive studies have been performed on the emergence of the
first objects that ended the dark age of the universe. Numerical
calculations adopting the cold dark matter models predicted
the reionization epoch in $z\sim 6-12$ (e.g. Gnedin \& Ostriker
1997).

Around this epoch a broad absorption trough in the blue part of
Ly$\alpha$ is expected and regarded as a strong indicator of
the reionization of the universe, which was predicted by Gunn \& Peterson 
(1965) and independently  also by Scheuer (1965).
 With the advent of the Hubble Space Telescope and 8 meter class
telescopes there have been extensive searches for the
Gunn-Peterson trough in the spectra of high red shift objects.
Remarkable contributions are made by the Sloan Digital Sky Survey, 
from which a number of high red shift quasars with $z$
ranging from 4 to 6 have been found. According to the recent
report from the Keck spectroscopy of these high red shift quasars
(Becker et al. 2001), the flux level drop around Ly$\alpha$ is
much higher for the quasar with $z=6.28$ than those for other
quasars with $z<6$, which indicates that the reionization epoch
may be found at around $z\sim 6$.

The exact computation of the flux drop around Ly$\alpha$  requires
an accurate atomic physical estimation of the scattering cross
section. Recent theoretical works on the calculation of the
Gunn-Peterson trough were provided by 
Miralda-Escud\'e (1998) and Madau \& Rees (2000), who adopted
the formula that was introduced in a heuristic way using the second
order time-dependent perturbation theory by Peebles (1993). The
formula is derived based on the assumption that the hydrogen atom
is a two-level atom, in order to show the behavior of the
scattering cross section that is approximated by the Lorentzian
near resonance and yields $\omega^4$ dependence in the low energy
limit. As Peebles noted clearly in the text, due to the two-level
assumption, the formula provides an inaccurate proportionality
constant in the low energy limit, even though it correctly gives
the $\omega^4$ dependence.

The accurate cross section should be obtained from the second order
time-dependent perturbation theory treating the hydrogen atom as
an infinitely many-level atom including the continuum free states, which
is known as the Kramers-Heisenberg formula. 
The discrepancy between the two formulae will be significant in the far
off-resonance regions where the contribution from the $np (n>2)$ states 
including the continuum states becomes considerable. Therefore, 
in order to obtain an accurate Gunn-Peterson profile
it is essential to investigate the exact scattering 
optical depth of a medium with a high neutral hydrogen column density.

In this Letter, we present a faithful atomic physics that governs
the scattering around Ly$\alpha$ by introducing the Kramers-Heisenberg 
formula and a simple fitting formula around Ly$\alpha$.  We compute the
Gunn-Peterson trough profiles adopting a representative set of
cosmological parameters with our choice of the reionization epoch
and make quantitative comparisons with previous works.

\section{The Kramers-Heisenberg Formula}

The interaction of photons and electrons is described by the
second order time-dependent perturbation theory, of which the
result is summarized as the famous Kramers-Heisenberg formula. As
is well illustrated in a typical quantum mechanics text, it may
be written as

\begin{eqnarray}
{d\sigma\over d\Omega}(\omega) &=& {r_0^2\over m_e^2\hbar^2}\Bigg\vert
\sum_{I}\left[ {\omega(\vec p\cdot
\hat\epsilon^{(\alpha')})_{AI}(\vec p\cdot
\hat\epsilon^{(\alpha)})_{IA}
\over
\omega_{IA}(\omega_{IA} -
\omega - i\Gamma_I/2)} \right. \nonumber \\
&-&\left. {\omega(\vec p\cdot
\hat\epsilon^{(\alpha)})_{AI}(\vec p\cdot
\hat\epsilon^{(\alpha')})_{IA}
\over
\omega_{IA}(\omega_{IA} +
\omega)}\right]\Bigg\vert^2
\end{eqnarray}
where $\hat\epsilon^{\alpha}, \hat\epsilon^{\alpha'}$ are the polarization
vectors associated with the incident photon and outgoing photon, respectively,
$\Gamma_I$ is the radiation damping term associated with the intermediate 
state $I$, $\omega$ is the incident angular frequency, $\omega_{IA}$ is
the angular frequency of the transition between I and the ground state $A$,
and $r_o={e^2/(m_e c^2)}= 2.82 \times 10^{-13}{\rm\ cm}$ is the
classical electron radius (e.g. Sakurai 1967). 

Here, the electron is in the ground
state $A$ before scattering and de-excites to the same state. The
summation (and integration) should be carried over all the
intermediate states $I$ including the infinite number of bound
states and the free or continuum states. 
For hydrogen, the dipole moment matrix elements have been explicitly
given using the recurrence relations of the hypergeometric
function in many texts and in the literature (e.g.
Berestetski, Lifshitz \& Pitaevskii 1971, Bethe \& Salpeter 1957, 
Karzas \& Latter 1961).

In the blue part of Ly$\alpha$, it is possible that the scattering
atom may de-excite to the excited $2s$ state by re-emitting a
photon with much lower frequency than the incident photon. This
inelastic scattering or the Raman scattering is negligible near
Ly$\alpha$ due to small phase space available for an outgoing
photon. However, this process becomes important as the incident
photon energy increases. In the range where the current work is
concerned, the Raman scattering process is safely neglected.

Despite the existence of the explicit analytic expressions of each
matrix element that constitutes the Kramers-Heisenberg formula for
hydrogen, it is still cumbersome to use the formula as it is.
Therefore, a simple fitting formula around Ly$\alpha$ will be 
useful for practical applications. Near resonance ($1170{\rm\
\AA}\ <\lambda<1410{\rm\ \AA}$), the Lorentzian function gives
quite a good approximation
\begin{equation}
\sigma(\omega)=\frac{3\lambda_\alpha^2}{8 \pi}
\frac{\Gamma_{2p}^2} {(\omega-\omega_\alpha)^2 + \Gamma_{2p}^2 /4},
\end{equation} 
where $\Gamma_{2p}=6.25\times10^8{\rm\ s^{-1}}$ is the radiation
damping constant associated with the Ly$\alpha$ transition.
In the long wavelength region
($\lambda>1410{\rm\ \AA}$), Gavrila (1967) provided the fitting
polynomial for the Rayleigh scattering cross section, which is
\begin{eqnarray}
\sigma(\omega)/\sigma_T &=&
0.400 (\omega/\omega_\alpha)^4 
+ 0.900 (\omega/\omega_\alpha)^6  \nonumber \\
&+& 12.6 (\omega/\omega_\alpha)^{14}.
\end{eqnarray}
In particular, Ferland (2001)
applied Gavrila's fit to his photoionization code `Cloudy'.

In the case of the short wavelength region 
($1070{\rm\ \AA}\ <\lambda<1170{\rm\ \AA}$), 
we provide a similar polynomial fit to
the Kramers-Heisenberg formula
\begin{eqnarray}
\sigma(\omega)/\sigma_T &=&1.62\times10^6 (\omega_\alpha/\omega)^4
+5.88\times10^6 (\omega_\alpha/\omega)^3 \nonumber \\
&+&7.99\times10^6 (\omega_\alpha/\omega)^2
-4.83\times10^6 (\omega_\alpha/\omega) \nonumber \\
&+&1.09\times10^6.
\end{eqnarray}
The deviation of the fit is within 5 per cent from the true
Kramers-Heisenberg formula. 

In Fig.~1 we show the scattering cross section from the
Kramers-Heisenberg formula by the solid line and by the dotted
line we represent the fit. The behavior near resonance is depicted in the
bottom panel, because the cross section changes very steeply. 
It is apparent that the scattering
cross section is excellently approximated by the Lorentzian. 
However, in the wavelength range considered in this Letter, 
the radiation damping is completely 
negligible, and the curve shown in the figure is simply proportional
to $\Delta\omega^{-2}=(\omega-\omega_{Ly\alpha})^{-2}$.  The deviation 
is slightly anti-symmetric with respect to the line center in the
sense that the cross section in the blue part is smaller than the Lorentzian
and in the red part it is larger than the Lorentzian. 

Therefore, 1 per
cent of deviation of the Lorentzian from the Kramers-Heisenberg formula
is seen at a wavelength shift
of $\Delta\lambda=\lambda-\lambda_\alpha = \pm 3.3{\rm\ \AA}$, for which
the corresponding cross section $\sigma =3.8\times 10^{-21}{\rm\ cm^2}$.
This indicates the accuracy of the Voigt profile fitting applied to
quasar absorption systems, where the
accuracy is more than 99 per cent when the absorbing medium is
characterized with the H~I column density smaller than $3\times 10^{20}
{\rm\ cm^{-2}}$.

Further away from
the line center, the  cross section in the blue part decreases very 
steeply till $\lambda\sim 1100 {\rm\ \AA}$, but the decrease of the cross
section in the red part is rather gradual and eventually becomes 
proportional to $\omega^4$, which corresponds to the classical result.

Fig.~1 also shows the comparison of the Kramsers-Heisenberg formula 
and the heuristic formula 
\begin{equation}
\sigma_P(\omega) = {3\lambda_\alpha^2\over 8\pi}
{\Gamma_{2p}^2 (\omega/\omega_\alpha)^4
\over
{(\omega-\omega_\alpha)^2+\Gamma_{2p}^2(\omega/\omega_\alpha)^6/4}}
\end{equation}
introduced by Peebles (1993).  

The $\omega^4$ dependence in the limit $\omega\ll\omega_\alpha$ 
is obtained as a result of the closure relation, which is apparent in the both 
formulae.  However,
the Kramers-Heisenberg formula gives about twice larger scattering
cross section than $\sigma_P$ does. This deviation is easily
noted when the oscillator strength of the Ly$\alpha$ transition is
$f_{Ly\alpha}=0.42$. The scattering cross section in the far red
region is contributed from all the $p$ states and the oscillator
strength is a good measure of the contributions of each individual
excited state. This implies that the $2p$ state contribution is
comparable to the total contributions from the remaining states in
the low energy limit.

\section{Gunn-Peterson Trough Profiles with the
Kramers-Heisenberg formula}

We compute the Gunn-Peterson optical depth defined by
\begin{equation}
\tau_{GP}=\int_{z_{rei}}^{z_s}\ dz\ {dl\over dz}\sigma[\nu=
c(1+z)/\lambda_{obs}]\ n(z),
\end{equation}
with the Kramers-Heisenberg formula and make comparisons with
previous works performed by Madau \& Rees (2000) (see also
Miralda-Escud\'e 1998). Here, $\lambda_{obs}$ is the observed wavelength,
$z_{rei},\ z_s$ are the redshifts of the complete reionization of the
universe and the reionizing source, and $n(z)=n_0(1+z)^3$ is the homogeneous
neutral hydrogen density at redshift $z$.  We choose
$z_{rei}=6,\ z_{s}=7$ as in Madau \& Rees (2000), but do not consider 
the proximity effect of the ionizing source. The Gunn-Peterson optical 
depth can be written as
\begin{eqnarray}
\tau_{GP} &=& N_{0\, HI} \int_{z_{rei}}^{z_s}\ dz\ 
\sigma[\nu=c(1+z)/\lambda_{obs}] \nonumber \\
&\times& (1+z)^2/[\Omega_M(1+z)^3 +\Omega_\Lambda]^{1/2},
\end{eqnarray}
where $\Omega_M, \Omega_\Lambda$ are the density parameters due to matter
and the cosmological constant and the characteristic hydrogen column density 
\begin{equation}
N_{0\, HI}\equiv n_0\ c\ H_0^{-1}.
\end{equation}
We choose the present Hubble constant and the hydrogen number density
$H_0=50{\rm\ km\ s^{-1}\ Mpc^{-1}},\ 
n_0=2.4\times10^{-7}{\rm\ cm^{-3}}$ so that $N_{0\, HI}=4.3\times
10^{21}{\rm\ cm^{-2}}$.

In Fig.~2, we show the Gunn-Peterson transmission coefficient $T_{GP}\equiv
e^{-\tau_{GP}}$.  We plot $T_{GP}$ for the case $\Omega_M=1, 
\Omega_\Lambda=0$ in the top panel and the same quantity 
for an accelerated expanding
universe $\Omega_M=0.35, \Omega_\Lambda=0.65$ in the bottom panel. 
In terms of the characteristic Gunn-Peterson
optical depth $\tau_{0\, GP}$ defined as
\begin{equation}
\tau_{0\, GP}(z_s)\equiv {3\lambda_\alpha^3\Gamma_{2p} n(z_s)
\over 8\pi H(z_s)},
\end{equation}
our choice of parameters in the case of the top panel yields 
the value of $\tau_{0\, GP}=3\times 10^5$ at $z_s=7$ as was adopted in the work 
of Madau \& Rees (2000).

The horizontal axis represents the logarithm of the normalized wavelength 
ratio $\delta$ defined as
\begin{equation}
\delta \equiv {\lambda_{obs}\over \lambda_\alpha(1+z_s)}-1,
\end{equation}
from which $\delta=0$ corresponds to the resonance wavelength of Ly$\alpha$.
By the dotted lines
we present the Gunn-Peterson transmission coefficient obtained using the
scattering cross section given by Eq.~(5). The deviation between the
two formulae is notable around $\delta =10^{-2}$, where 
the deviation is about 3 per cent in the top panel and 5 per cent 
in the bottom panel. The Peebles approximation turns out to be pretty 
good for contemporary application.

The deviation between the two formulae will increase as $n_0$ or
$N_{0\, HI}$ increases, because the discrepancy of the Kramers-Heisenberg 
formula and Eq.~(5) becomes larger as the frequency is further away from
the line center. Near resonance, both formulae are excellently approximated
by the same Lorentzian. Therefore, no significant deviation is expected
when the neutral medium is of low column density $\la 10^{21}{\rm\ cm^{-2}}$.
It is notable that an accurate treatment of atomic physics is more important
in an accelerated expanding universe where the
univere was more compact than the universe without the cosmological constant.

\section{Summary and Discussion}

In this Letter, we have investigated the behavior of the scattering
cross section around Ly$\alpha$ in a quantitative way, where the deviation 
from the Lorentzian becomes significant as the incident frequency gets further
away from the line center. Therefore, in an analysis of the Gunn-Peterson
trough profile, which is associated with a neutral medium with a high H~I
column density, an inaccurate treatment of
the atomic physics of hydrogen may introduce significant errors in estimating
important cosmological parameters including the epochs of the emergence
of the first objects and the completion of the reionization of the universe.

Voigt profile fitting has been very successfully applied to
quasar absorption systems with a broad range of H~I column densities. 
However, the deviation of the true scattering cross
section from the Lorentzian exceeds 1 per cent
when the relevant column density becomes $N_{HI}\ge 3\times
10^{20} {\rm\ cm^{-2}}$ that is the typical column density of a
damped Ly$\alpha$ absorber.  This is especially
important in some damped Ly$\alpha$ systems that may possess $N_{HI} >
10^{21}{\rm\ cm^{-2}}$ (e.g. Turnshek \& Rao 1998). However, it should
be noted that the damping constant $\Gamma_{2p}$ is so small compared
with the scale relevant in this work, the cross section is effectively of the
form $\propto\Delta\omega^{-2}=(\omega-\omega_\alpha)^{-2}$. Therefore, the
absorption profile is irrelevant to the exact value of the 
radiation damping term, which means that the term `damped Ly$\alpha$ 
absorption' is a misnomer.

\acknowledgments
KHB thanks the support from the BK21 project
initiated by the Ministry of Education. HWL thanks Roger Blandford and
Sang-Hyeon Ahn for their helpful discussions.

\clearpage

\begin{figure}
\plotone{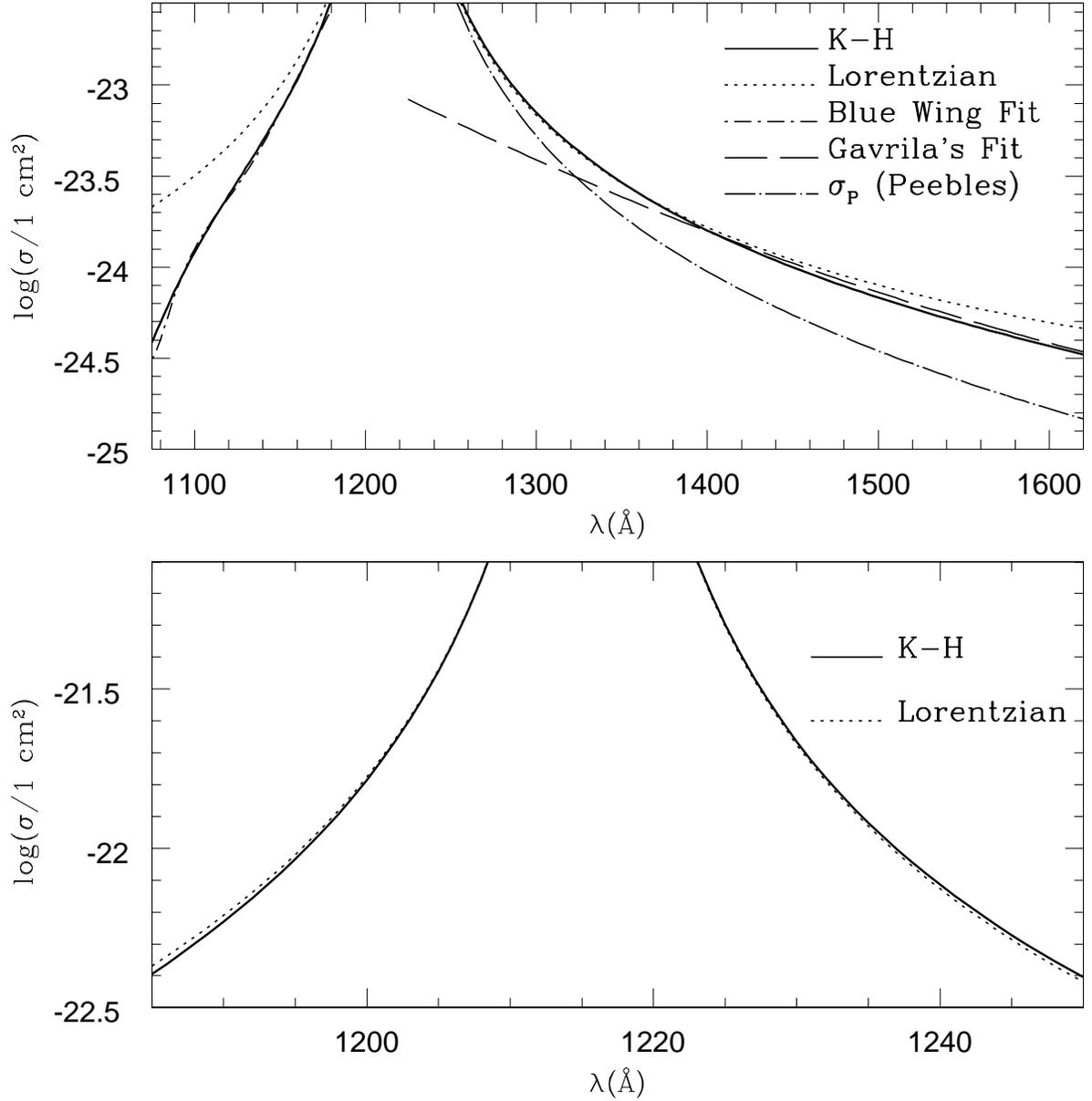} \caption{
The scattering cross section around Ly$\alpha$. The solid line represents
the accurate, fully quantum mechanical cross section known as the 
Kramers-Heisenberg formula. The dotted line represents the Lorentzian
given by Eq.~(2), which gives an excellent approximation near the line
center. The long dashed line represents the fitting formula provided by
Gavrila (1967), of which the approximation is valid for $\lambda>1400{\rm
\ \AA}$. The dot-dash line represents our fit to the Kramers-Heisenberg
formula in the blue part given in Eq.~(4). The dot-long dash line represents
the cross section obtained from Eq.~(5),
which is inaccurate by a factor of two in the far red wing region
and accurate near the line center. In the bottom panel, we plot the cross
section obtained from the Kramers-Heisenberg formula (solid line) and 
the Lorentzian Eq.~(2).  The cross section is asymmetric relative 
to the line center in the sense that the cross section in the blue part 
is smaller than in the red part.
\label{fig1}}
\end{figure}

\clearpage

\begin{figure}
\plotone{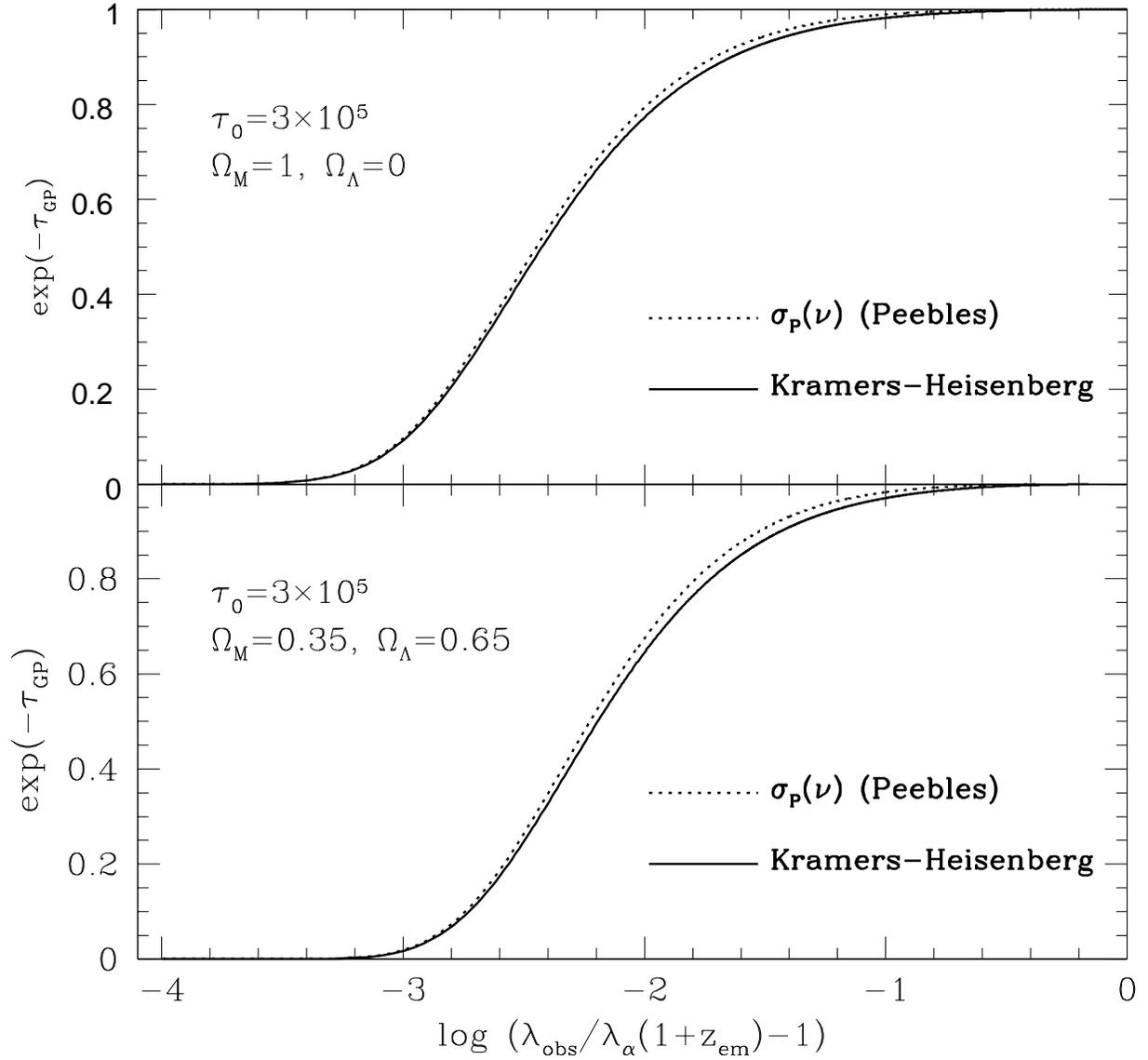} \caption{
The Gunn-Peterson transmission coefficient $T_{GP}\equiv e^{-\tau_{GP}}$
for $\Omega_M=1, \Omega_\Lambda=0$ (top panel) and for
$\Omega_M=0.35, \Omega_\Lambda=0.65$ (bottom panel). 
The solid lines represent the values obtained using the Kramers-Heisenberg
formual and the dotted lines are for the values from the cross section
$\sigma_P$ introduced by Peebles (1993).
The present Hubble constant and the hydrogen number density are chosen to be
$H_0=50{\rm\ km\ s^{-1}\ Mpc^{-1}},
n_0=2.4\times10^{-7}{\rm\ cm^{-3}}$ so that $N_{0\, HI}=n_0 c H_0^{-1}=
4.3\times 10^{21}{\rm\ cm^{-2}}$.
\label{fig2}}
\end{figure}

\end{document}